\documentclass[]{aa}

\usepackage[varg]{txfonts}
\usepackage{graphicx}
\usepackage{natbib}
\usepackage{amsmath}
\usepackage{booktabs}
\usepackage{xspace}

\begin{document}

\title{A giant magnetic eruption from the asymptotic giant branch star Mira A}

\author{
W.~Vlemmings\inst{1}\thanks{Corresponding author:
wouter.vlemmings@chalmers.se}
\and
M.~Andriantsaralaza\inst{1}
\and
B.~Bojnordi Arbab\inst{1}
\and
T.~Khouri\inst{1}
\and
B.~Lankhaar\inst{2}
\and
L.~Planquart\inst{1}
\and
E.~De Beck\inst{1}
\and
S.~Del Palacio\inst{1}
\and
E.~Humphreys\inst{3}
\and
D.~Jadlovsky\inst{4, 3}
\and
M.~Maercker\inst{1}
\and
H.~Olofsson\inst{1}
\and
M.~Saberi\inst{2,5}
\and
M.~Siebert\inst{1}
\and
R.~Unnikrishnan\inst{1}
\and
S.~Wedemeyer\inst{2,5}
\and
M.~Wittkowski\inst{3}
}

\institute{
Department of Physics and Astronomy,
Chalmers University of Technology,
41296 Gothenburg, Sweden
\and
Institute of Theoretical Astrophysics,
University of Oslo,
PO Box 1029 Blindern,
0315 Oslo, Norway
\and
European Southern Observatory (ESO),
Karl-Schwarzschild-Stra{\ss}e 2,
85748 Garching bei M{\"u}nchen, Germany
\and
Department of Theoretical Physics and Astrophysics, 
Faculty of Science, 
Masaryk University, Kotl{\'a}\u{r}sk{\'a} 2, 61137, Brno, Czech Republic
\and
Rosseland Centre for Solar Physics,
University of Oslo,
PO Box 1029 Blindern,
0315 Oslo, Norway
}

\abstract {Mass loss during the advanced evolutionary stages of low-
  and intermediate-mass stars is traditionally attributed to
  dust-driven winds, whereas the influence of magnetic fields remains
  poorly constrained. However, recent observations show massive
  directional mass ejections from the nearby asymptotic giant branch
  (AGB) star Mira~A that appear not to be driven by radiation pressure on the dust.}  {We
  observationally investigate whether magnetic fields are responsible
  for the mass loss ejections from Mira A.}  {We obtained
  full-polarisation ALMA observations of the SiO $v=1$,
  $J=5\rightarrow4$ maser transition at 215~GHz with angular
  resolutions of 1.5--2.6~au.}  {The observations reveal a highly
  magnetized loop-like mass ejection traced by strongly polarised SiO
  masers extending to approximately eight stellar radii. The
  morphology, polarisation pattern, energetics, and kinematics
  indicate a magnetically driven eruption analogous to a stellar
  coronal mass ejection.}  {These observations demonstrate that
  magnetic eruptions can dominate mass loss from evolved stars and
  imply that magnetic fields must become an integral component of
  future models of AGB, and potentially other cool giant, winds.}

\maketitle

%======================================================================
\section{Introduction}
%======================================================================

In the current picture of mass loss from asymptotic giant branch (AGB)
stars, pulsations and convection lift material to a sufficient
distance from the star for it to cool and form dust.  Subsequently,
radiation pressure transfers momentum to the small dust particles,
which in turn drag along the gas and produce a steady stellar wind when averaged on many-year timescales \citep{Hofner2018}.  This
mass-loss process is crucial for the chemical enrichment of the
Universe.

Since strong magnetic fields have been detected throughout the
circumstellar envelopes of AGB stars as well as on and near the
stellar surface \citep[e.g.][]{Herpin2006,Lebre2014,Vlemmings2014}, it
has been suggested that magnetic fields may play a role in the
generation of AGB mass loss \citep[e.g.][]{Yasuda2019}.  However, the
effects of magnetic fields are not included in state-of-the-art
wind-driving models \citep[e.g.][]{Freytag2017}, and magnetic fields
have not been empirically shown to cause mass loss.

Recent observations of oxygen-rich (M-type) AGB stars have challenged
the current wind-driving scenario by revealing individual AGB stars
with apparent regular mass loss but with insufficient dust to drive
the winds \citep{Schirmer2025}.  Additionally, several phenomena that
cannot be explained by the current mass-loss paradigm have been
observed toward nearby AGB stars. This includes strong hotspots
\citep[e.g.][]{Vlemmings2017} and directional mass ejections
\citep[e.g.][]{Hoai2023}. Specifically the archetypal M-type AGB star
Mira A, also known as omicron Ceti, stands out as a source with many
unexplained features, including hot spots on the (sub-)millimeter surface
and in its atmosphere \citep{Andriantsaralaza2026}, large asymmetric mass
ejections \citep{Khouri2026}, and X-ray flares \citep{Karovska2005}.

To investigate the underlying mechanism producing these observed
features, we observed Mira A in full-polarisation mode with the
Atacama Large Millimeter/submillimeter Array (ALMA). The observations,
data reduction and source properties are described in
\S~\ref{obs}. The resulting maser distribution and polarisation
detections are presented in \S~\ref{results} and in \S~\ref{discussion}
we describe the maser polarisation interpretation, derive magnetic
field lower limits and ejection energetics and discuss the possible
origin of the strong magnetic field. Finally in \S~\ref{conclusion} we
conclude how the observations can introduce a paradigm shift in our
understanding of evolved star mass loss.

\section{Observations}
\label{obs}

The AGB star Mira A and its companion Mira B were observed in ALMA
Band~6 \citep{Ediss2004} as part of the joint VLA and ALMA project
2024.1.01776.S (PI: Vlemmings). The observations were obtained on 2025
August 10 in full polarisation mode using four spectral windows (spws)
centred on $214.065,\; 215.573,\; 228.483,\; {\rm and}\; 231.176~{\rm
  GHz}.$ The 214.065 and 215.573 GHz spectral windows were centred on
the SiO $v=1$ and $v=2$, $J=5\rightarrow4$ transitions, respectively,
with rest frequencies of $215.59595~{\rm GHz}$ and $214.08854~{\rm
  GHz}$ \citep{Muller2001}. These spectral windows had bandwidths of
58.594 MHz and 960 channels, yielding a channel spacing of
approximately $85~{\rm m\,s^{-1}}.$

The remaining spectral windows each had a bandwidth of $ 1.875~{\rm
  GHz} $ with 960 channels and a corresponding channel width of $
2.73~{\rm km\,s^{-1}}.  $ The integration time of the individual
visibilities was $ 2.02~{\rm s}.  $ The observations were obtained in
the ALMA C--9 configuration, with baselines ranging from $ 164~{\rm m}
$ to $ 14\,670~{\rm m}.  $ The maximum recoverable scale was
approximately $ 0\farcs4.  $ The quasars J2258$-$2758 and J0217+0144
were used as bandpass/amplitude and phase calibrators, respectively,
while J0006$-$0623 served as polarisation calibrator.  The total
observing time was 3.5 hours, of which approximately 70 minutes were
spent on source.  Pipeline-calibrated data products were delivered
through the ESO ALMA Regional Centre. Calibration was performed using
the ALMA pipeline implemented in CASA version 6.6.6.17
\citep{Hunter2023,CASATeam2022}.  After the initial calibration,
including polarisation calibration, the data were further processed
locally using CASA version 6.6.1.  First, molecular lines were
identified and flagged before the data were averaged to an integration
time of $ 6.06~{\rm s} $ and to 50 channels per spectral window.  Two
rounds of phase-only self-calibration were then performed on the
continuum dataset. This improved the continuum signal-to-noise ratio
by approximately a factor of $ 3.7.  $ The resulting self-calibration
solutions were subsequently applied to the un-averaged line dataset
after restoring the flagged spectral channels.  The spectral window
containing the SiO $v=1$ maser transition was transformed from the
TOPO reference frame to the LSRK frame.  An additional phase-only
self-calibration step was then performed on the brightest maser
channel.  For the line data, this increased the signal-to-noise ratio
of the strongest channel by approximately an additional factor of $ 8.  $ The
resulting calibration table was applied to the full line dataset.  A
continuum image was produced using all line-free channels from the
four spectral windows and \texttt{superuniform} visibility weighting.
The resulting synthesized beam size was $ 16 \times 15~{\rm mas}, $
with a position angle of $ 59.3^\circ.  $ We additionally imaged all
four Stokes parameters using Briggs weighting (\texttt{robust}=0.5),
producing a beam of $ 26 \times 20~{\rm mas} $ with a position angle
of $ 55.6^\circ.  $ No continuum linear or circular polarisation was
detected.  The $3\sigma$ rms sensitivity was $
46.8~\mu{\rm Jy\,beam^{-1}}, $ corresponding to fractional linear and
circular polarisation limits of approximately $ 0.1\%.  $ The SiO
$v=1$, $J=5\rightarrow4$ maser emission was then imaged using Briggs
weighting with \texttt{robust}=0.5 at a spectral resolution of $
0.1~{\rm km\,s^{-1}}.$ The synthesized beam for the maser cubes was $
28 \times 21~{\rm mas} $ with a position angle of $ 55.4^\circ.  $
Imaging was performed in Stokes $I$, $Q$, $U$, and $V$.  The SiO
$v=2$, $J=5\rightarrow4$ transition was not detected.  To extract
individual maser components we used the \texttt{tclean} component map,
representing the deconvolution of the data into point-source
components. Subsequently, a clustering analysis was performed using a
spatial link length of $ 3~{\rm mas} $ and a velocity link scale
of $ 1~{\rm km\,s^{-1}}$. To identify a maser as a single
feature we required at least seven CLEAN components within the
clustering scale and present in at least three consecutive velocity
channels.

\begin{figure}
\centering \includegraphics[width=\columnwidth]{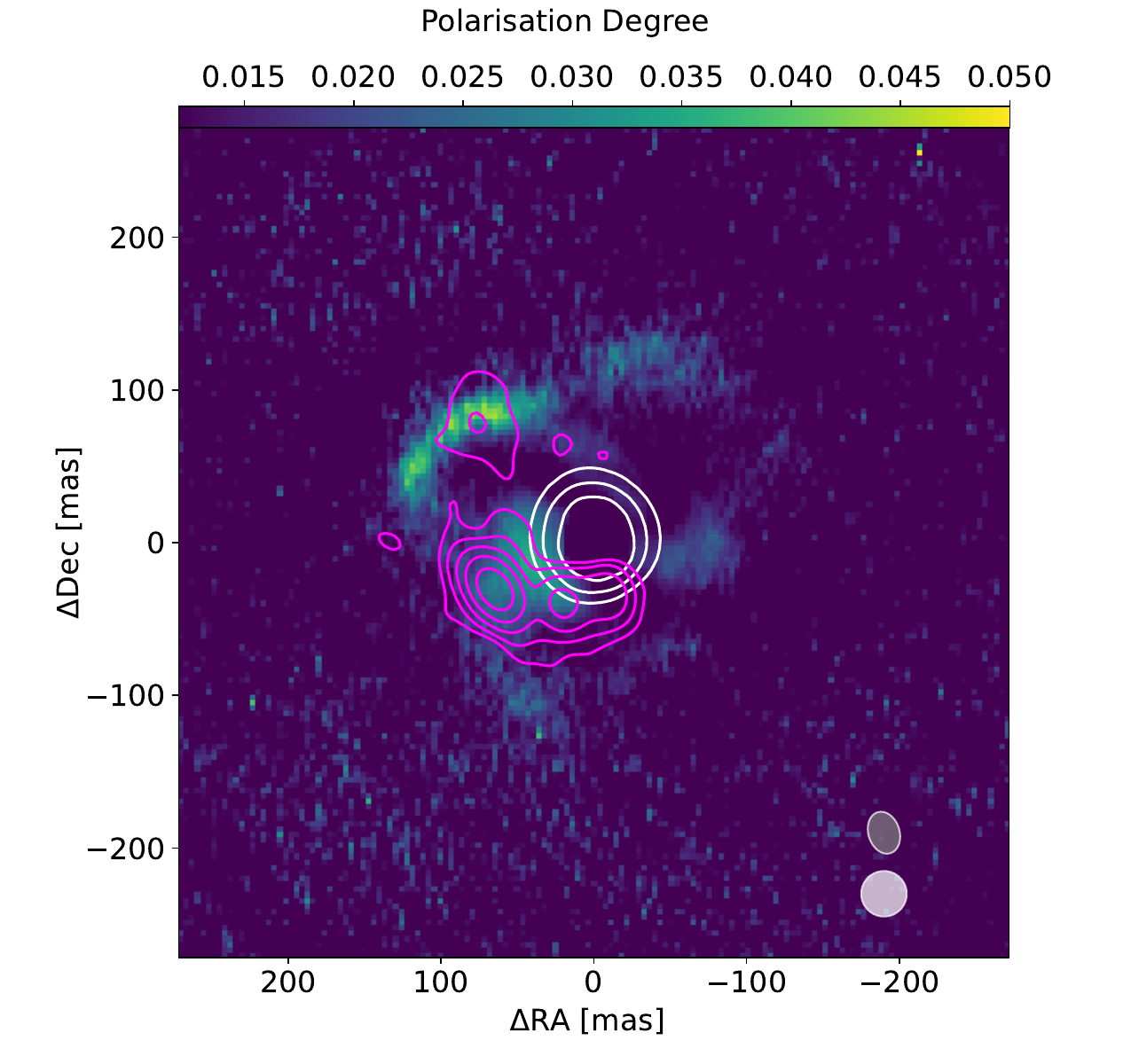}
\caption{ Comparison between the SiO $v=1$, $J=5\rightarrow4$ masers
  observed with ALMA during Science Verification in 2014 and
  VLT/SPHERE dust-polarisation map obtained in September 2015
  \citep{Khouri2026}.  Magenta contours indicate integrated SiO
  emission at levels of $0.04, 0.08, 0.16, 0.32,$ and $0.64$ times the
  peak integrated intensity of $5.58~${\rm
    Jy\,beam$^{-1}$\,km\,s$^{-1}$}. White contours show the ALMA
  stellar continuum emission. The colour scale represents the degree
  of polarised dust emission observed in the 644.9\,nm (CntHa) filter. The ellipses in the bottom right represent the ALMA beam (top) and the Sphere PSF FWHM (bottom).
}
\label{fig:MiraSphereSiO}
\end{figure}

\begin{figure*}
\centering
\includegraphics[width=\textwidth]{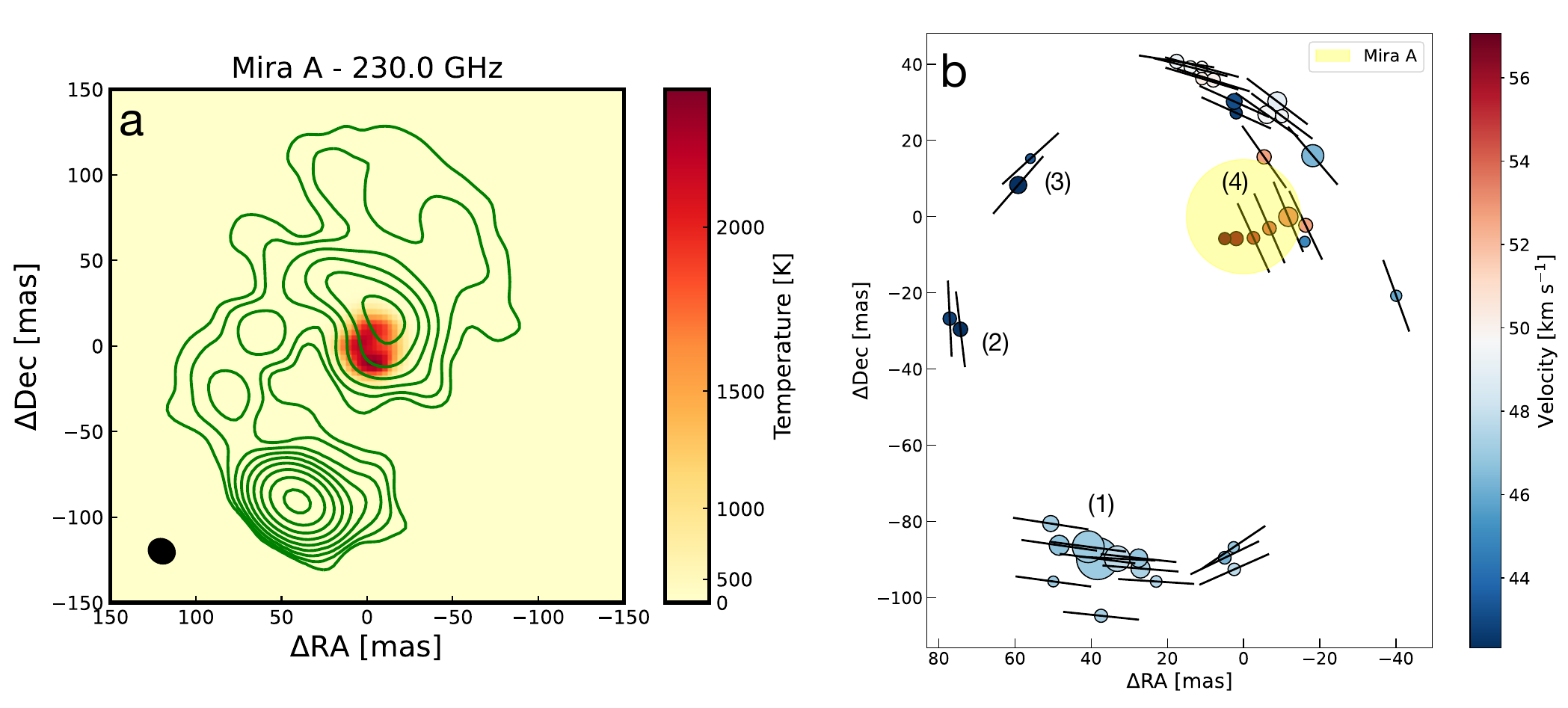}
\caption{ SiO maser emission towards Mira A observed in 2025.  (a) Moment 0 map of the
  SiO $v=1$, $J=5\rightarrow4$ emission around Mira A.  The green
  contours are the integrated SiO emission at levels of 0.8, 0.4, 0.2,
  \ldots, 0.00325 times the maximum ($212.6~{\rm
    Jy\,beam^{-1}\,km\,s^{-1}}$).  The minimum contour corresponds to
  approximately $7\sigma$, with an rms noise of $0.1~{\rm
    Jy\,beam^{-1}\,km\,s^{-1}}$.  The colour scale represents the
  brightness temperature of the 230-GHz stellar surface with a peak of
  2340 K.  The synthesized beam is shown in the lower left corner.
  (b) Maser features around Mira A, with the infrared photosphere
  shown in yellow. The stellar velocity of Mira A $V_{\rm lsr}\approx47.7~{\rm
    km\,s^{-1}}$. Features are colour coded by LSR velocity and
  scaled by flux.  Black line segments indicate EVPAs for masers with
  linear polarisation detected above $5\sigma$ ($13~{\rm
    mJy\,beam^{-1}}$).  Numbers identify the maser clusters displayed
  in Fig.~\ref{fig:spectra}. }
\label{fig:masers}
\end{figure*}

\begin{figure*}
\centering
\includegraphics[width=\textwidth]{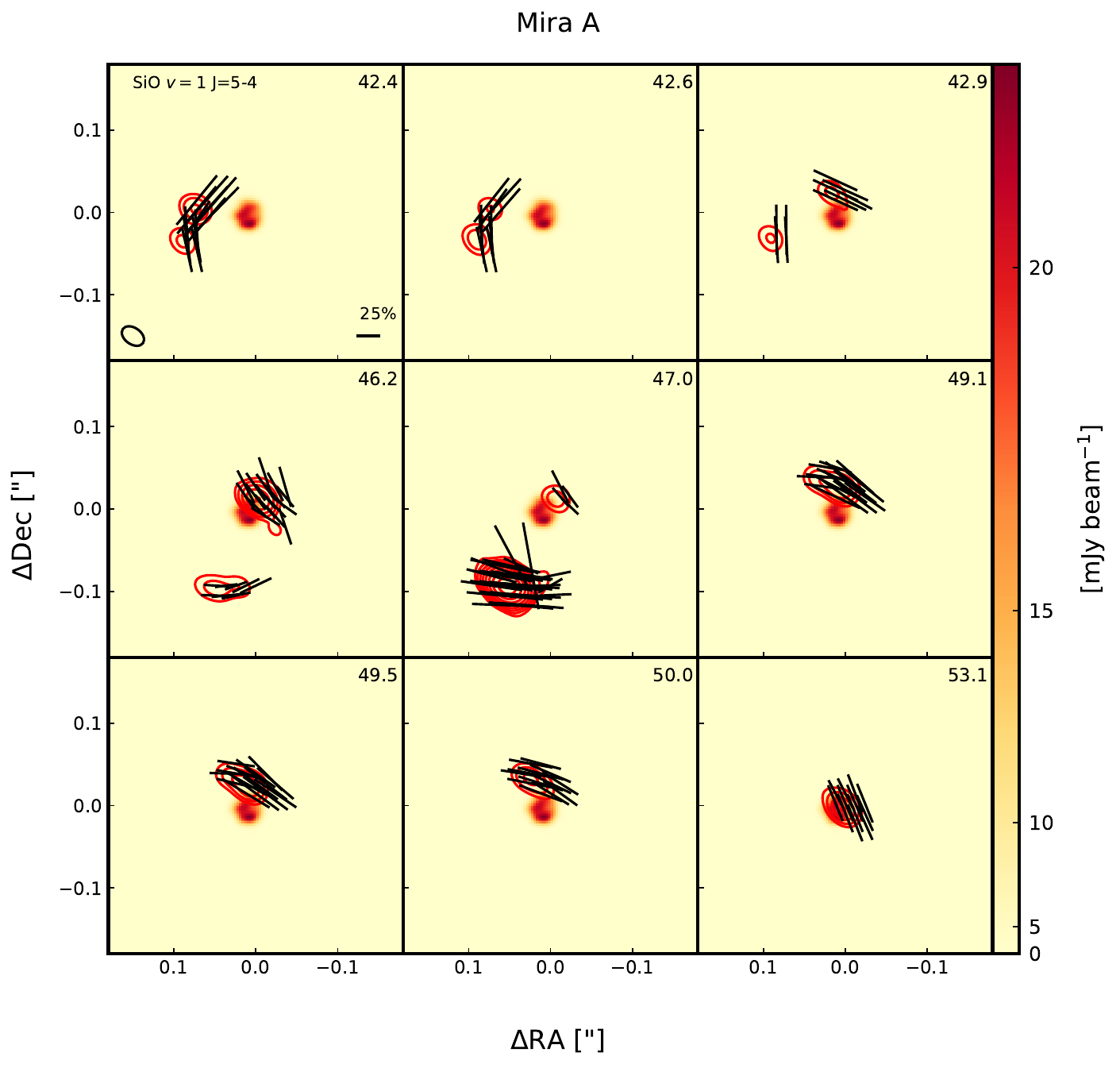}
\caption{Selected velocity-channel maps of the SiO $v=1$,
  $J=5\rightarrow4$ maser emission around Mira A. Contours show the
  maser emission at 0.1 km\,s$^{-1}$ spectral resolution, superposed
  on the stellar continuum map reconstructed using superuniform
  weighting. The channel velocity is indicated in the upper-right
  corner of each panel. The displayed channels correspond to the peak
  emission of individual maser features and are not equally spaced in
  velocity. Contours are drawn at $0.64, 0.32, 0.16, \ldots, 0.005$
  times the peak maser emission of $245 \pm 12$~ {\rm
    Jy\,beam$^{-1}$}. The rms noise in line-free channels is $13$~{\rm
    mJy\,beam$^{-1}$}. Line segments indicate Nyquist-sampled EVPAs
  where linear polarisation is detected at greater than
  $5\sigma$. Vectors are scaled according to fractional polarisation
  as indicated in the upper-left panel. Also shown in that panel is
  the synthesized beam of the Briggs-weighted maser cube.  }
\label{fig:channels}
\end{figure*}

%======================================================================

\begin{figure*}
\centering
\includegraphics[width=\textwidth]{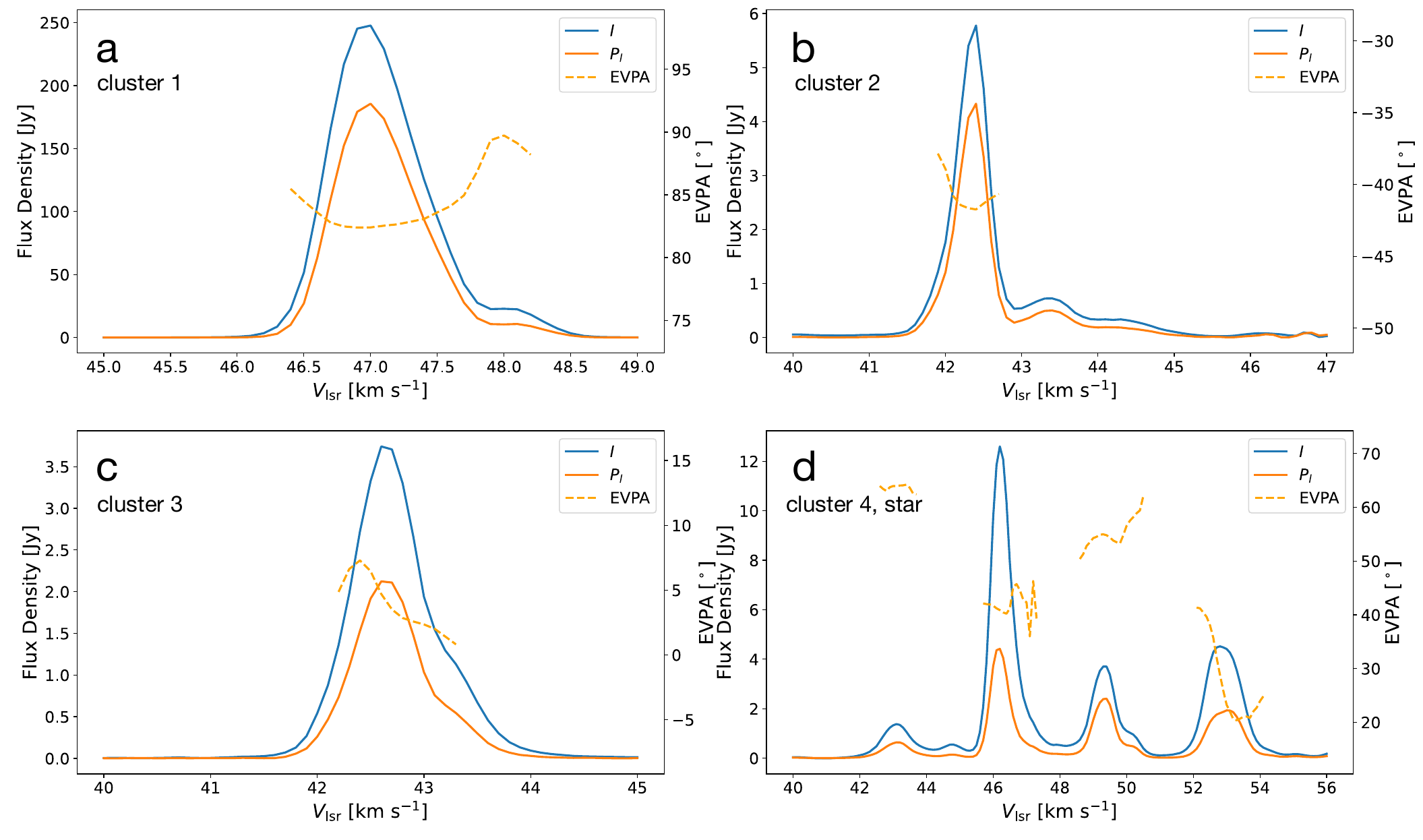}
\caption{Stokes $I$ and linear-polarisation spectra for the four
  strongest SiO maser peaks around Mira A. Panels (1)--(4) correspond
  to the four maser clusters identified clockwise from the brightest
  southeastern maser feature shown in Fig.~\ref{fig:masers}. The blue
  curve shows Stokes $I$, the orange curve shows the
  linear-polarisation intensity, and the dashed curve indicates the
  EVPA across the maser spectrum.  }
\label{fig:spectra}
\end{figure*}

%======================================================================

%======================================================================
\subsection{Mira A}
\label{mira}
%======================================================================

For consistency with recent literature, we adopt a distance of $ d\approx
100~{\rm pc} $ for Mira A \citep{vanLeeuwen2007}.  Mira A is the
archetype of the class of Mira-type AGB pulsators.  Its companion,
Mira B, has a projected separation of approximately $ 0\farcs4, $ and
has been identified as a white dwarf based on the detection of optical
flickering \citep{Sokoloski2010}.  Although X-ray emission has been
detected from Mira B, an X-ray outburst observed in 2003 was
determined to originate from the AGB primary rather than the compact
companion \citep{Karovska2005}.

The mass-loss rate of Mira A has primarily been estimated from
single-dish CO observations and is typically found to be a few times $
10^{-7}\ M_\odot\,{\rm yr^{-1}} $
\citep[e.g.][]{Planesas1990,DeBeck2010}.  This value remains uncertain
because of the complexity of the circumstellar environment.  For
example, the high-velocity outflow from Mira B interacts with the
slower molecular gas expelled by Mira A, producing a cavity
approximately $ 10'' $ in extent \citep{Ramstedt2014}.

During the past decade, high-angular-resolution observations of Mira A
at visible, millimetre, and submillimetre wavelengths have revealed
several unexpected phenomena related to the stellar atmosphere and
mass-loss process.  ALMA images of the stellar continuum show compact
hotspots with brightness temperatures exceeding $ 8000~{\rm K}, $
substantially above the average stellar brightness temperature of
approximately $ 2000~{\rm K} $ \citep{Vlemmings2019Atmospheres}.
Monitoring of these structures showed that they are more compact and
significantly hotter during optical maximum light, whereas during
minimum light they appear to be more extended and their brightness-temperature excess decreases to several hundred kelvin above that of the surrounding stellar surface \citep{Andriantsaralaza2026}.

Observations of circumstellar molecular gas indicate a dense extended
atmosphere surrounding Mira A. This region extends outward to
approximately $ 4\,R_\star $ and contains roughly $ 10^{-4}\ M_\odot $
of gravitationally bound gas \citep{Khouri2018InnerEnvelope}.
Based on measurements by \citet{Woodruff2009}, we adopt the smallest measured
infrared radius of Mira A as the photospheric radius $ R_\star = 15~{\rm mas} \simeq 1.5~{\rm au}.  $ 
  
Cooling within the extended atmosphere of Mira A appears highly
efficient, with temperatures decreasing to approximately $ 500~{\rm K}
$ at four $R_*$ \citep{Khouri2018InnerEnvelope}. Beyond the dense inner atmosphere, the
molecular gas distribution becomes increasingly complex owing to 
interaction with the companion and episodic asymmetric ejections from
Mira A itself. In particular, two large lobes containing approximately
$ 2\times10^{-5}\ M_\odot $ of gas and extending to approximately $
30~{\rm au} $ from the star have been detected using both visible
polarised light and molecular-line observations \citep{Khouri2026}.
The current expansion velocities in these structures range between $ 10 $
and $ 13~{\rm km\,s^{-1}}$. Polarised visible-light emission is
produced primarily at the outer boundaries of the lobes, indicating
that dust is predominantly concentrated in these regions.  ALMA Science Verification observations (2011.0.00014.SV) acquired a short time after the estimated ejection time and described in \citet{Humphreys2018} show that
the SiO $v=1$,
$J=5\rightarrow4$ masers trace the same ejection
(see Fig.~\ref{fig:MiraSphereSiO}).  The extent of
  the maser emission matches the expansion of the dust-scattering
  structure as well as later molecular-line observations
  \citep{Khouri2026}.

\begin{figure}
\centering
\includegraphics[width=\columnwidth]{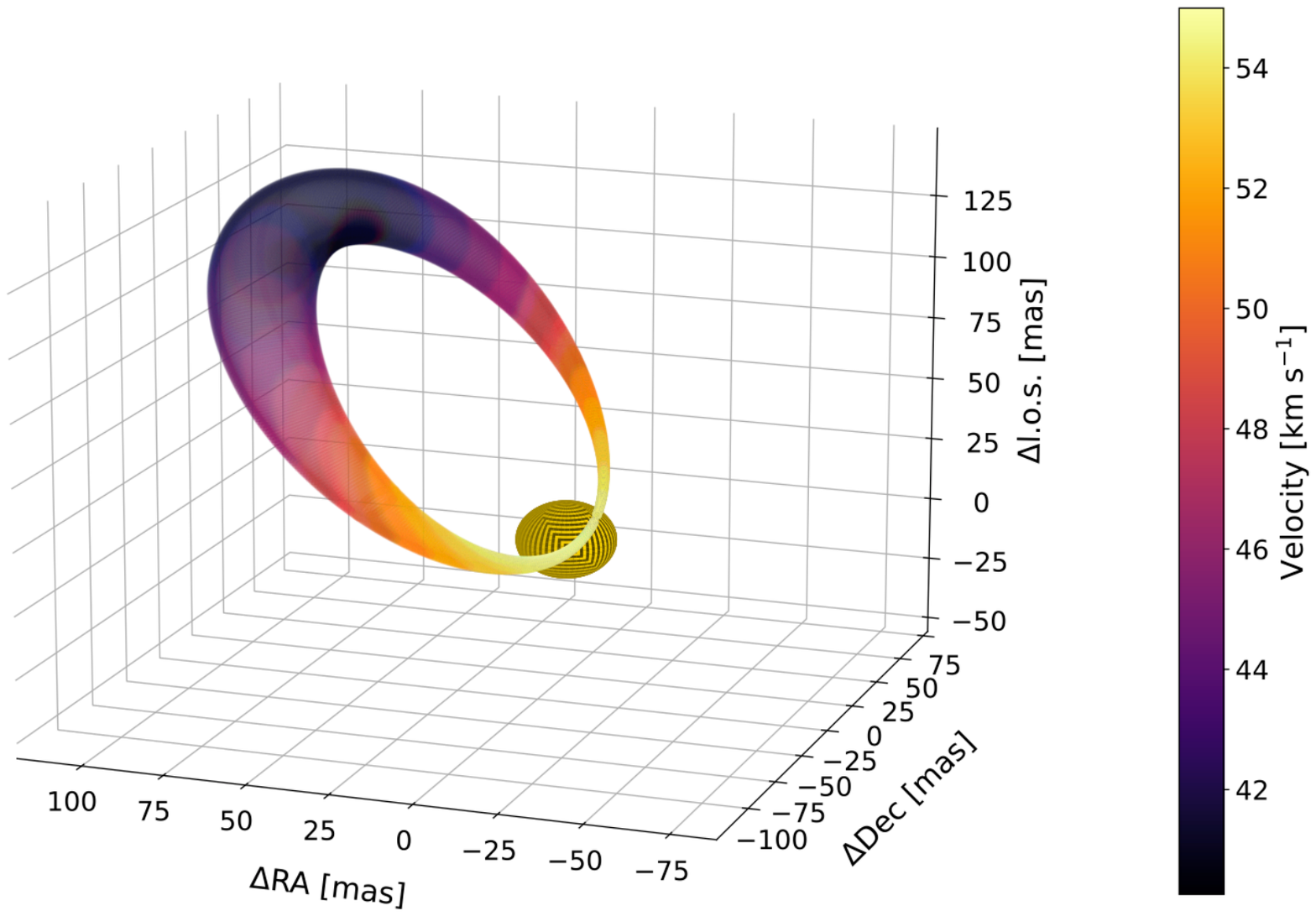}
\caption{ Three-dimensional representation of the adopted
  magnetic-loop ejection model anchored to the stellar surface. The
  structure is colour coded according to line-of-sight velocity. The observers view is along the axis labelled as $\Delta$l.o.s. from the top of the plot.}
\label{fig:3dloop}
\end{figure}

%======================================================================
\section{Results}
\label{results}
%======================================================================

We identified numerous linearly polarised SiO maser features in the new ALMA images, as well
as more extended emission reaching $120$ mas or 12 au from Mira A, or approximately
eight stellar radii (Fig.~\ref{fig:masers}a).  The masers are strongly
linearly and circularly polarised (Fig.~\ref{fig:masers}b and
\S~\ref{siomasers}) and display a loop-like, elliptical
distribution, with the brightest maser located at the apex of the
structure toward the southeast of the star. The electric vector
polarisation angles (EVPAs) closely trace the elliptical maser
structure. The positions, velocities, peak fluxes, fractional linear polarisation $(p_L)$ and EVPAs of the individual maser features are presented in Table~\ref{spots}. The uncertainty on the center velocity is less than the channel width of 0.1~km~s$^{-1}$ and the flux uncertainty is dominated by the absolute flux calibration uncertainty of $\sim5\%$.

An integrated intensity (moment~0) map of
the maser emission with respect to the stellar continuum is presented
in Fig.~\ref{fig:masers}a, while a component map of the masers and
their EVPAs is shown in Fig.~\ref{fig:masers}b.  The positional
uncertainty of individual maser features depends on signal-to-noise
ratio and image reconstruction quality.  For the masers displayed in
this work, the uncertainty ranges from approximately $ 0.8~{\rm mas} $
to $ 3.5~{\rm mas}.$ Channel maps of the strongest maser features, together with their EVPA
vectors and the stellar continuum emission, are shown in
Fig.~\ref{fig:channels}.  Spectra of the brightest maser clusters, together
with a spectrum for the line of sight toward the stellar disc, are shown in
Fig.~\ref{fig:spectra}, where we display Stokes $I$, the
de-biased linear polarisation ($P_{\rm l}$, which is a function of the two Stokes parameters $Q$ and $U$ as well as the rms noise $\sigma_{Q,U}$ in the polarisation maps)
\begin{equation}
P_{\rm l} = \sqrt{Q^2 + U^2 - \sigma_{Q,U}^2},
\end{equation}
and the EVPA.

The SiO maser distribution exhibits a predominantly elliptical
structure extending toward the south-east, together with weaker
emission in the northeastern quadrant. To describe the distribution of
the strongest maser features, we adopt an elliptical loop anchored to
the stellar surface. This representation is intended as a simplified
description of the observed morphology and resembles geometries
commonly associated with halo and light-bulb coronal mass ejections (CMEs),
Kopp--Pneuman magnetic loops, kink-instability flux tubes, or
flux-rope structures discussed in the CME literature \citep{Webb2012}. Given the diversity of stellar
eruptions \citep{Webb2012}, the turbulent nature of the extended
atmosphere, the likely presence of remnants from previous eruptions,
and the strong dependence of maser emission on favourable
amplification paths, the proposed loop should be regarded as an
idealized representation rather than a unique physical solution. 

The adopted ad hoc loop geometry consists of a tube whose radius increases
from $ 0.015~{\rm au} $ at the stellar surface to $ 1.5~{\rm au} $ at
the apex.  The tube is wrapped around an ellipse with dimensions $
15~{\rm au} $ along the major axis and $ 14~{\rm au} $ along the minor
axis.  The thickness of the tube is chosen to reproduce an inferred
maser amplification path length of approximately $ 3~{\rm au} $ near
the apex of the structure.  Toward the stellar surface the tube
thickness decreases as $ \propto r^{-2}$, where $r$ is the distance to the star The apex of the loop has a
projected position angle of $ 95^\circ.  $ The loop is assumed to
expand ballistically following a launch velocity of approximately $
35~{\rm km\,s^{-1}}.  $ This velocity vector is directed toward the
apex of the loop and is inclined by $ 73^\circ $ with respect to the
plane of the sky.  The loop is anchored to the south-west quadrant of
the stellar surface, with a footpoint size of approximately $ 0.5~{\rm
  au}.  $ A three-dimensional visualization of this geometry is
presented in Fig.~\ref{fig:3dloop}, and a parametric description of the loop is presented in Appendix~\ref{loop-model}. In
Fig.~\ref{fig:loopmodel} we show the maser feature distribution with
the loop model as seen projected on the plane of the sky as well as in
an RA-velocity projection. As noted, the majority of the masers trace
the loop structure, even if several masers also trace the more familiar
SiO maser zone located within approximately $ 2\text{--}3\,R_\star $
of the star. Nevertheless, the magnetic, kinetic and thermal
energies inferred in \S~\ref{energy} remain of comparable magnitude
even if the true three-dimensional morphology differs from the adopted
loop geometry.

\begin{figure*}
\centering
\includegraphics[width=\textwidth]{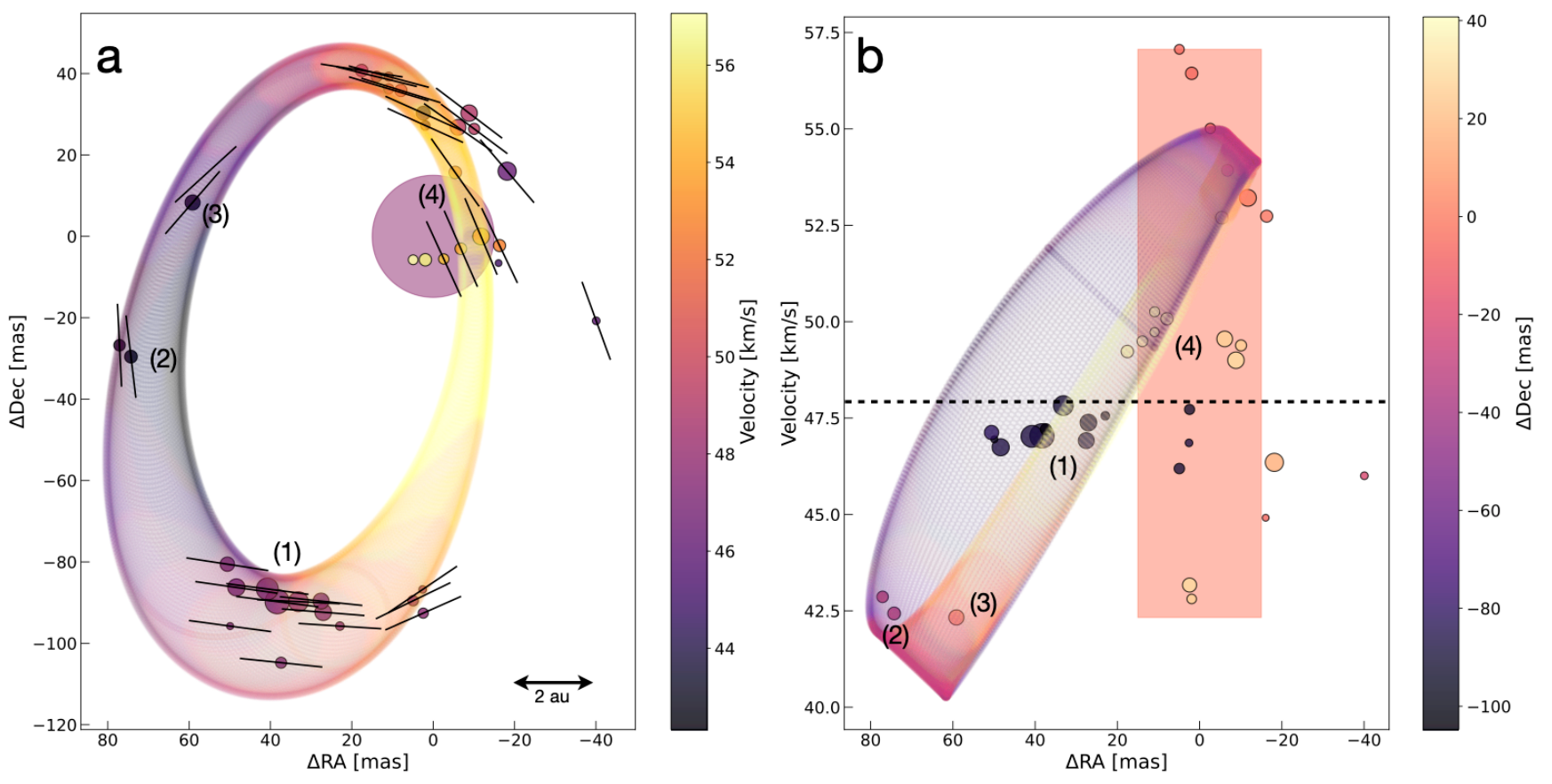}
\caption{ Illustration of the magnetic loop model that matches the
  observations.  (a) Same as Fig.~\ref{fig:masers}b with the fitted
  loop overlaid. The loop is colour coded by LSR velocity along the
  line of sight. The large circle denotes Mira A and is colour coded
  according to its stellar velocity $V_{\rm lsr}\approx47.7~{\rm
    km\,s^{-1}}$.  (b) Model in the RA--velocity plane.  The
  rectangular region indicates the star and the velocity range
  associated with the extended atmosphere.  The dashed horizontal line
  marks the stellar velocity.  The loop reaches an expansion velocity
  of $16.4~{\rm km\,s^{-1}}$ in the direction of ejection, corresponding
  to a maximum ballistic launch velocity of approximately $35~{\rm
    km\,s^{-1}}$.  }
\label{fig:loopmodel}
\end{figure*}

%==================

\section{Discussion}
\label{discussion}

\subsection{SiO maser polarisation interpretation}
\label{siomasers}

Various levels of polarisation, both circular and linear, have been
observed toward SiO, H$_2$O, and OH masers in circumstellar envelopes
\citep[e.g.][]{Herpin2006,Vlemmings2014}.  The interpretation of maser
polarisation generally depends on the degree of saturation, possible
anisotropic pumping, and the magnetic field along the amplification
path \citep[e.g.][]{Lankhaar2019,Nedoluha1994}.  To interpret the
polarisation measurements we need to establish the relative magnitudes
of the maser stimulated-emission rate $R$, the decay rate $\Gamma$,
and the magnetic precession rate $g\Omega$, as these determine whether
the polarisation traces the magnetic field.  The stimulated-emission
rate can be written as
\begin{equation}
R = A_{ij} \frac{k_{\rm B}}{h\nu} 
\frac{T_{\rm b} \Delta\Omega}{4\pi} \quad {\rm s}^{-1},
\end{equation}
where $A_{ij}$ is the Einstein coefficient of the maser transition,
$\nu$ is the transition frequency, $T_{\rm b}\Delta\Omega$ is the
emerging maser brightness temperature, and $k_{\rm B}$ and $h$ are the
Boltzmann and Planck constants, respectively.  For the SiO $v=1$,
$J=5\rightarrow4$ transition, $ A_{ij} = 5.134\times10^{-4} ~{\rm
  s^{-1}}, $ and $ \nu = 215.5959~{\rm GHz}.  $ The emerging
brightness temperature can be estimated from the observed maser
properties assuming a beaming angle $\Delta\Omega$.  For circumstellar
SiO masers, $ \Delta\Omega \approx 10^{-1} $ is typically adopted
\citep{Elitzur1992}.  The observed brightness temperature is
\begin{equation}
T_{\rm b} = 6.13\times10^{11} \left( \frac{S_\nu}{\rm Jy} \right)
\left( \frac{\nu}{\rm GHz} \right)^{-2} \left( \frac{\theta}{\rm mas}
\right)^{-2} {\rm K},
\end{equation}
where $S_\nu$ is the flux density of the maser and $\theta$ is the angular size
of the maser.  For the brightest maser feature we measure $ S_\nu =
191\pm9~{\rm Jy}, $ with an angular size of approximately $ \theta =
4~{\rm mas}, $ corresponding to diameter $ D \approx 0.4~{\rm au}.  $ This
yields $ T_{\rm b} = (1.6\pm0.1)\times10^8 ~{\rm K}.  $ For the weaker
polarised masers, $ T_{\rm b} > 10^5~{\rm K}.  $ Assuming a
characteristic maser path length of $ l \approx 1~{\rm au}, $ typical
for circumstellar SiO masers, the beaming angle becomes
\begin{equation}
\Delta\Omega = \left( \frac{D}{l} \right)^2 = 0.16.
\end{equation}
We thus derive a stimulated-emission rate for the brightest maser of $
R \approx 100~{\rm s^{-1}}, $ while weaker masers have values one to
three orders of magnitude smaller.  If the true maser size is smaller,
$T_{\rm b}$ increases while $\Delta\Omega$ decreases such that $R$
remains approximately constant.

Masers become saturated when the stimulated-emission rate exceeds the
decay rate.  For SiO masers in the first vibrational state, $ \Gamma
\approx 5~{\rm s^{-1}} $ \citep{Elitzur1992}, implying that several of
the masers observed here are likely saturated.  The large fractional
linear polarisation supports this interpretation because saturated
masers generally produce higher linear polarisation fractions than
unsaturated masers. However, saturation alone is not sufficient to
explain the observed polarisation levels.  Anisotropic pumping is
required to reach the polarisation fractions of up to approximately $
75\% $ measured in the spectra shown in Fig.~\ref{fig:spectra}
\citep{Lankhaar2024}.

\begin{figure}
\centering
\includegraphics[width=\columnwidth]{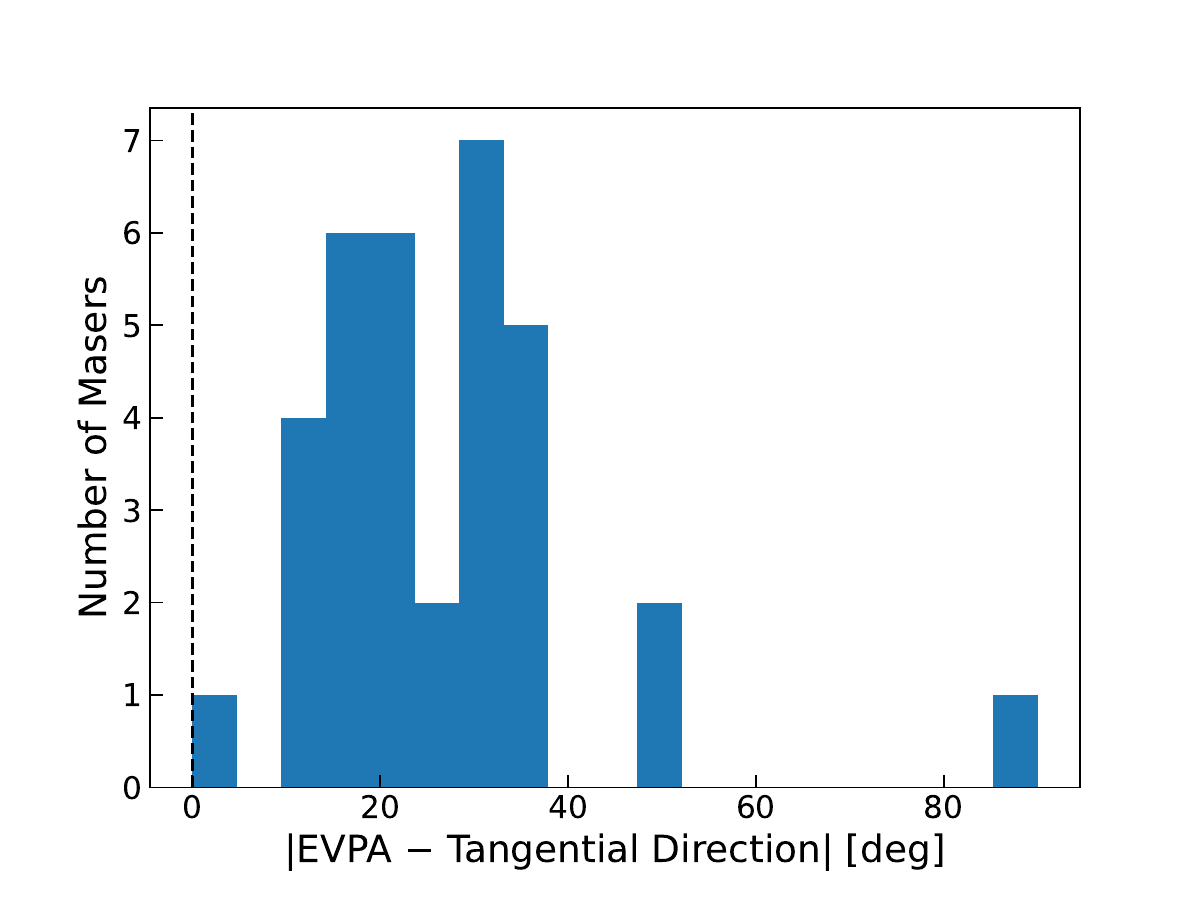}
\caption{Distribution of the absolute value of the EVPA offset from
  the local tangential direction at the position of each maser. The
  distribution demonstrates that the EVPAs are neither systematically
  radial nor tangential with respect to the central star.}
\label{fig:evpa}
\end{figure}

The observed polarisation direction traces the magnetic-field
direction only if the magnetic field defines the symmetry axis of the
masing molecules. This condition is fulfilled when $ g\Omega \gg R.  $
If the stimulated-emission rate approaches the magnetic precession
rate, the molecular symmetry axis becomes aligned with the radiation
field rather than with the magnetic field, producing significant
changes in polarisation angle \citep{Lankhaar2019,Lankhaar2024}.
Because velocity redistribution is slow compared with
stimulated-emission events, we expect $ R \propto S_\nu $ within
individual spectral channels.  If $ R \sim g\Omega, $ significant
variations in EVPA would therefore be expected across the line
profile.  Instead, we observe that the EVPAs remain stable to within
only a few degrees across the spectral line for almost all masers.
The small observed variations are likely caused by blending of
individual maser components within the synthesized beam as well as, for the channels that include the strongest maser, dynamic range effects.
Additionally, as shown in Fig.~\ref{fig:evpa}, the EVPAs do not show
systematic radial or tangential alignment relative to the central
star. Together, these results indicate that neither anisotropic
pumping nor maser amplification geometry controls the observed
polarisation direction.  We therefore conclude that the masers are
operating in the regime $ g\Omega \gg R > \Gamma .  $ In this limit
magnetic precession dominates over stimulated emission, and the
linear-polarisation vectors align either parallel or perpendicular to
the magnetic-field direction, depending on the angle between the
magnetic field and the line of sight. The strong anisotropic pumping
inferred from the high linear polarisation together with the
predominantly tangential EVPA distribution suggest that the
polarisation vectors are mainly parallel to the magnetic field
\citep{Lankhaar2024}. Because the polarisation morphology follows the
maser structure, which we interpret as an outward-propagating shock,
the magnetic field would naturally lie along the shock front.

Magnetic-field components parallel to a shock are compressed, whereas
perpendicular components are not.  We therefore conclude that the
linear polarisation vectors trace the magnetic-field direction within
the maser-emitting gas.  For SiO masers, the magnetic precession rate
is $ g\Omega \simeq 1.5\, B_{\rm mG} \ {\rm s^{-1}}, $ where $B_{\rm
  mG}$ is the magnetic-field strength expressed in
milligauss. Adopting the conservative criterion $ g\Omega \gtrsim 10R,
$ to ensure magnetic dominance over stimulated emission, we find $ B
\gtrsim 0.7~{\rm G}.  $ This represents a robust lower limit to the
magnetic-field strength in the SiO maser loop.  

\begin{figure}
\centering
\includegraphics[width=\columnwidth]{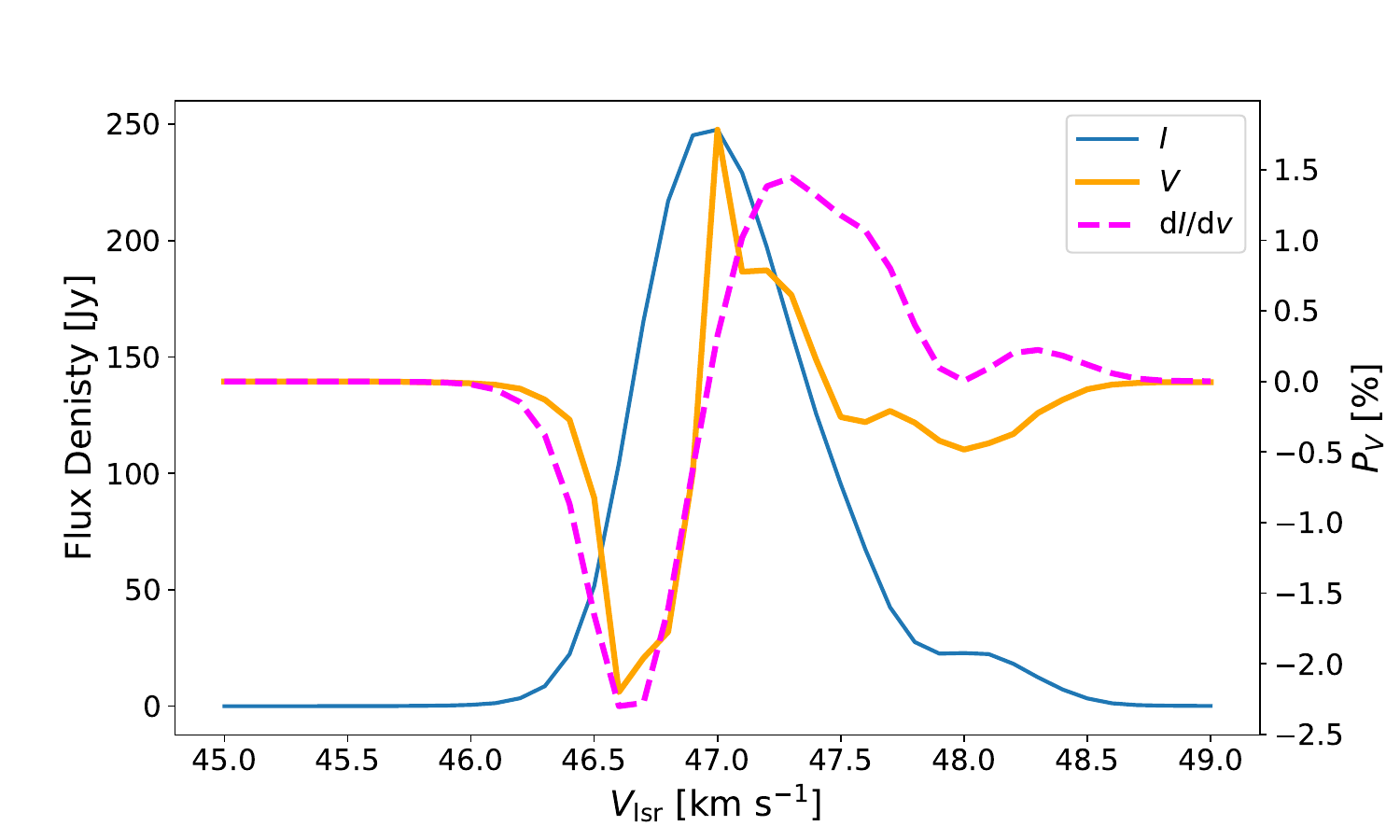}
\caption{ Stokes $I$ and circular-polarisation spectra of the
  strongest SiO maser feature. The blue curve shows the Stokes-$I$
  intensity and the orange curve the circular-polarisation fraction
  $V$ scaled to the peak emission. The dashed curve represents the
  derivative of the Stokes-$I$ spectrum scaled to correspond to a
  magnetic-field strength of approximately 80~G under a pure Zeeman
  interpretation. }
\label{fig:vspec}
\end{figure}

The brightest maser
additionally exhibits significant circular polarisation, shown in
Fig.~\ref{fig:vspec}. Under a standard Zeeman interpretation, the
circular polarisation spectrum displays the familiar S-shaped profile
proportional to the velocity derivative of the Stokes-$I$ spectrum.
The proportionality constant is determined by the Zeeman splitting
coefficient. Neglecting saturation and non-Zeeman effects, the
observed circular polarisation corresponds to a line-of-sight magnetic
field strength of approximately $ B_{\parallel} \approx 80~{\rm G}.  $
For the weaker masers, similar calculations yield magnetic-field
strengths of several tens of gauss.  However, non-Zeeman effects can
significantly enhance circular polarisation in saturated and
anisotropically pumped masers \citep{Lankhaar2019,Wiebe1998}.
Consequently, the circular polarisation likely overestimates the true
magnetic-field strength. We therefore adopt the result derived from
the linear polarisation analysis and conclude that the magnetic-field
strength in the maser loop is at least $ B \gtrsim 0.7~{\rm G}.  $
%======================================================================

%======================================================================
\subsection{Physical characteristics of the maser loop}
\label{phys}

The pumping of $v=1$ SiO masers requires high densities, $n_{\mathrm{H}_2}
\sim 10^{9}-10^{10}\ {\rm cm^{-3}}$, with the masers quenched above
this range, and the strongest masers additionally require a sufficient
SiO column density along a velocity-coherent path
\citep[e.g.][]{Lockett1992,Bujarrabal1981}. Consequently, SiO masers
are typically observed in tangential amplification zones within two to
three stellar radii of the star \citep[e.g.][]{Desmurs2014}, with
models indicating that the extent of the masers decreases for
higher-$J$ transitions \citep{Lockett1992}.

The brightest maser feature in our observations is located at
approximately nine stellar radii, where the ambient gas density is
expected to be one to two orders of magnitude below that required for
strong SiO maser emission \citep{Ireland2011}.  We therefore infer
that the masers trace an outward-propagating shock in which the gas
density is substantially enhanced. From the measured dimensions of the
SiO maser region, with a major axis of 15 au, a minor axis of 14 au,
and an assumed thickness of approximately 3 au at the apex
(Fig.~\ref{fig:loopmodel}), we estimate an ejection
volume of $V \approx 3\times10^{41}~{\rm cm^3}$.

Given the presence of very strong SiO $v=1$, $J=5\rightarrow4$ masers,
we adopt a conservative lower limit on the average gas density,
$\langle n_{\mathrm{H}_2} \rangle > 2\times10^{9}~{\rm cm^{-3}}$.
This implies a total ejected mass of $M_{\rm ej} \gtrsim
1.1\times10^{-6}\ M_\odot$. Based on the loop geometry and velocity,
approximately 30\% of the ejected material does not reach escape
velocity $(v_{\rm esc} =
30\mbox{--}37~{\rm km\,s^{-1}}$, assuming a stellar mass of
$0.8$--$1.2\,M_\odot$.).  Hence
it remains gravitationally bound and will experience fallback.

The velocity of the masers indicates a ballistic launch velocity of
approximately $v_{\rm launch} \approx 35~{\rm km\,s^{-1}}$. We
estimate the Alfv\'en velocity in the magnetized ejection using
$v_{\rm A} = \frac{B} {\sqrt{4\pi\rho}}$, where $\rho$ is the average
mass density, and find $v_{\rm A} \gtrsim 23~{\rm km\,s^{-1}}$.  If we
assume that the magnetic field is frozen into the gas and that
  the gas expands isotropically while moving outwards , $B \propto
  {\rho}^{2/3}$. Hence, in that case, the Alfv\'en velocity $v_{\rm
    A}\propto \rho^{1/6}$. For a density at launch that is roughly two
  orders of magnetic higher than the current density in the maser
  loop, the launch Alfv\'en velocity is approximately 1.5-2.0 times
  higher than the surface escape velocity from Mira A or the launch
  velocity inferred from the observations. This indicates that
  magnetic tension was sufficient to overcome gravitational
  confinement, driving the observed eruption.

\begin{figure}
\centering
\includegraphics[width=\columnwidth]{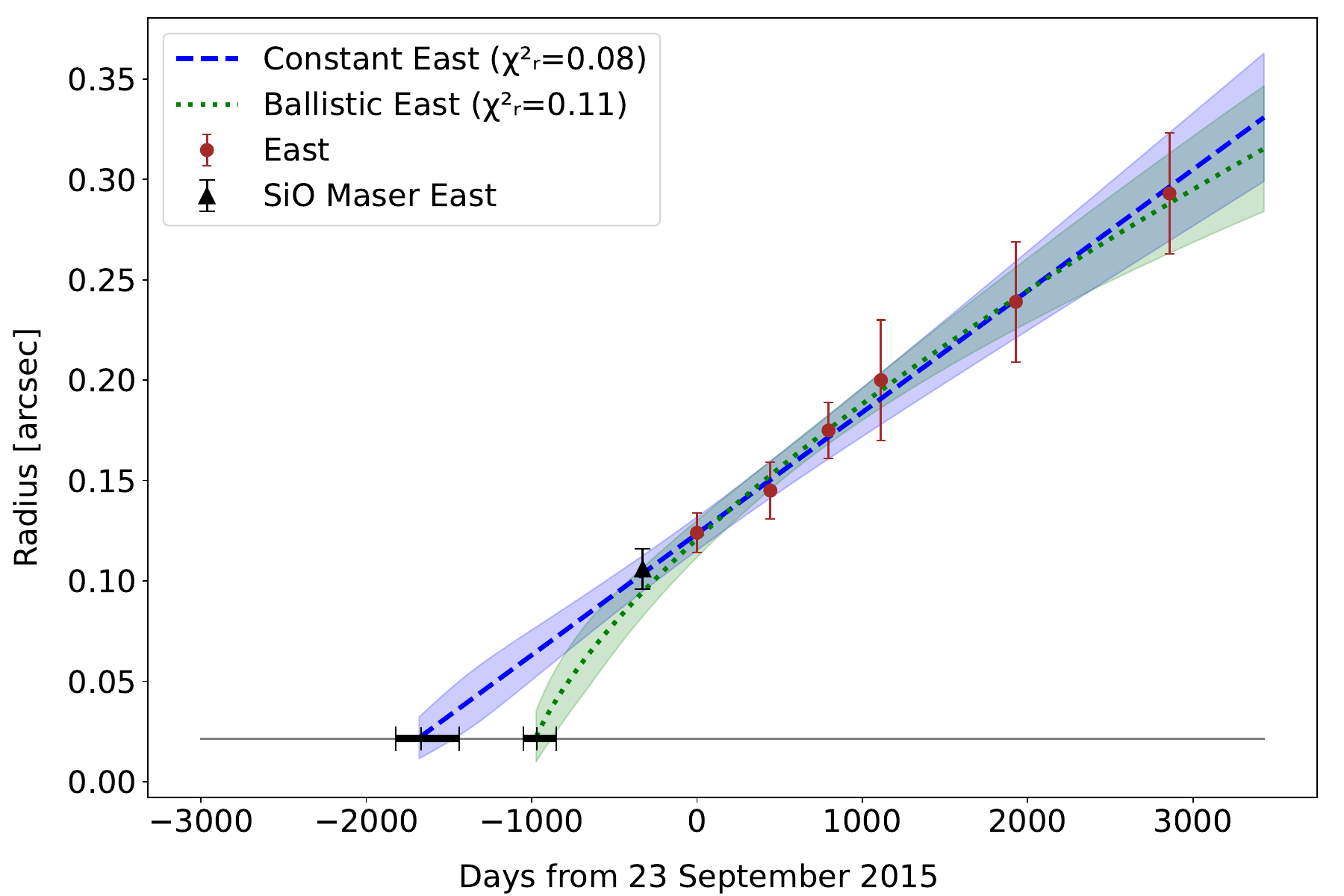}
\caption{ Expansion history of the eastern ejection traced by dust and
  molecular-line observations \citep{Khouri2026}. The measured extent
  of the SiO $v=1$, $J=5\rightarrow4$ maser emission shown in
  Fig.~\ref{fig:MiraSphereSiO} is plotted as the leftmost point, demonstrating
  that the SiO masers trace the onset of the directional mass
  ejection. The dashed blue line represents a constant-velocity
  expansion model, while the dotted green line indicates a ballistic
  trajectory launched close to the stellar escape velocity. The
  horizontal black bars indicate the inferred launch intervals for the
  two kinematic descriptions.  }
\label{fig:expansion}
\end{figure}

Observations of the SiO masers around Mira A in 2014 revealed a
similar aspherical maser distribution toward the east of the star
\citep{Humphreys2018}. As shown in Fig.~\ref{fig:MiraSphereSiO}, these
masers trace the edge of a ballistically expanding mass ejection with
a mass of $\sim10^{-5}\,M_\odot$, detected in dust and molecular
emission \citep{Khouri2026}. In Fig.~\ref{fig:expansion}, we update
figure 3 from \cite{Khouri2026} with the SiO maser position that
indicate that they match the expected ballistic ejection. That event,
together with a second ejection toward the northwest, was launched in
late 2010 or early 2011.  No polarisation measurements are available
for the masers observed in 2014.  Given the strong morphological
similarities with the previous ejection, whose onset was traced by the
SiO $v=1$, $J=5\rightarrow4$ masers, we suggest that the ejection
observed in the 2025 ALMA data presented here represents another episode of
highly directional mass loss from Mira A.

Assuming a ballistic ejection launched at approximately the escape
velocity, the latest event would have originated between 1.3 and 2
years prior to our recent observations, that is, between mid-2023 and
early 2024. This timing is consistent with the detection of a hotspot
in June 2023 on the southwest quadrant of the submillimeter surface,
with a brightness temperature exceeding 4200 K above the submillimeter
surface temperature \citep{Andriantsaralaza2026}.

%======================================================================

\subsection{Energy estimates}
\label{energy}

We estimate the magnetic and kinetic energies of the ejection by
adopting the observed morphology, average expansion velocity, and an
ejection mass of $M_{\rm ej} \sim 1.1\times10^{-6}\,M_\odot$ (derived in \S~\ref{phys}).

The kinetic energy is given by
\begin{equation}
E_{\rm K} = \frac{1}{2} M_{\rm ej} \langle v_{\rm ej}\rangle^2 .
\end{equation}
Assuming a launch velocity of approximately $ v_{\rm ej} \approx
35~{\rm km\,s^{-1}}$, we obtain $ E_{\rm K} = 1.4\times10^{40} ~{\rm
  erg}$.

The magnetic energy contained within the ejection is
\begin{equation}
E_{\rm B} = \left( \frac{B^2}{8\pi} \right)V,
\end{equation}
where $V$ is the full volume of the ejection. For $B > 0.7~{\rm G}$,
this implies $E_{\rm B} > 0.6\times10^{40} ~{\rm erg}$.

The magnetic and kinetic energies are therefore comparable, indicating
that the magnetic field is dynamically significant and capable of
organizing and channeling the mass ejection.  The total energy of the
ejection exceeds the energies typically observed in stellar
superflares \citep{Notsu2019} by up to several orders of magnitude.

We also compare the ejection energy to the thermal energy of the
hotspot detected on the submillimeter surface of Mira A in
June 2023.

The thermal energy is given by
\begin{equation}
E_{\rm th} = \frac{3}{2} n k_{\rm B} T V_{\rm spot},
\end{equation}
where $n$ is the gas number density at the submillimeter surface,
$k_{\rm B}$ is the Boltzmann constant, $T$ is the spot temperature,
and $V_{\rm spot}$ is the spot volume. Using $n \approx 10^{12}~{\rm
  cm^{-3}}$, $T > 4200~{\rm K}$, a spot size smaller than
$1.9\times0.7~{\rm au}^2$, and a characteristic vertical scale of
approximately $0.8~{\rm au},$ yields $E_{\rm th} > 1.0\times10^{40}
~{\rm erg}.$ The thermal energy is therefore of the same order of
magnitude as the magnetic and kinetic energies, suggesting that the
energy released during the outburst both heats and accelerates the
gas.

Finally, we can show that the result is not dependent on potential large uncertainties in the derived volume of the ejection, by comparing the local energy densities of the ejecta and hotspot. The current
magnetic energy density is:
\begin{equation}
u_{\rm B} = \frac{B^2}{8\pi}.
\end{equation}
For $B\gtrsim0.7$~G we find $u_{\rm B}\approx2\times10^{-2}~{\rm erg\,cm^{-3}}$.

We evaluate the kinetic energy density using maximum loop velocity ($v_{\rm ej}\approx16.4~{\rm km\,s^{-1}}$) derived from the observations and the
density adopted for the SiO maser ($n_{\rm ej}\approx2\times10^{9}~{\rm cm^{-3}}$ . The current kinetic energy density is
\begin{equation}
u_{\rm K} = \frac{1}{2}\rho_{\rm ej}\,v_{\rm ej}^2
\approx 9\times10^{-3}~{\rm erg\,cm^{-3}}.
\end{equation}
This is comparable to, but somewhat below, the magnetic energy density
($u_{\rm K}/u_{\rm B}\approx0.5$).

The thermal energy density of the hotspot at the launch of the ejecta in June 2023 is,
\begin{equation}
u_{\rm th} = \frac{3}{2} n k_{\rm B} T.
\end{equation}
This means that $u_{\rm th}\gtrsim0.9~{\rm erg\,cm^{-3}}$ for the density and temperature values adopted previously.

Hence, both the kinetic and magnetic energy are roughly two orders of
magnitude below the thermal energy density of the hotspot. As shown in
\S.~\ref{phys}, the ejecta was close to magnetic--kinetic
equipartition at launch, while the magnetic energy density is slightly
larger than the kinetic energy density in the more diluted gas in
August 2025.  This is consistent with the magnetic field diluting more
slowly than the kinetic energy density as the ejecta expands and
decelerates. As expected, the thermal energy density at the hotspot
remains substantially higher since at launch the energy was
concentrated in a much smaller and denser region.

%======================================================================
\subsection{Magnetic field origin and implications}
%======================================================================

The origin of the magnetic fields responsible for the observed
ejections remains uncertain, although several mechanisms have been
proposed \citep{KonstantinovaAntova2025}. A turbulent dynamo operating
within the deep convective envelope can provide a baseline magnetic
field \citep[e.g.][]{Soker2002}. In addition, differential rotation
between the stellar core and envelope offers a natural source of shear
capable of amplifying and organizing the field on larger spatial
scales \citep[e.g.][]{Blackman2001,Nordhaus2008}. In that case, as
mass loss proceeds and the envelope mass decreases, magnetic fields of
fixed strength would become progressively more dynamically
important. The envelope evolution may therefore lower the threshold
for magnetic-flux emergence and eruptive events. Alternatively, the
engulfment of a companion or massive planet could enhance dynamo
action \citep{Livio2002}.  Such an interaction would likely impart
significant angular momentum to the convective envelope, potentially
inducing measurable rotation. Rotation has been detected in the AGB
star R~Doradus \citep{Vlemmings2018Rotation}, but no comparable
rotational signature is observed in Mira~A.

The observations presented here together with the dust and gas observations from 
\citet{Khouri2026} reveal a class of recurrent, localized mass
ejections whose morphology, kinematics, and magnetic structure are
difficult to reconcile with standard wind-driving mechanisms
considered for AGB stars. The ejections appear to originate from
confined surface regions, exhibit coherent loop-like magnetic
geometries extending several stellar radii, and are launched
impulsively at velocities comparable to the local escape speed,
followed by almost ballistic propagation. Dust formation appears to
occur only at the outer boundaries of the ejecta \citep{Khouri2026},
indicating that dust is formed because of the mass ejections and that
it is not the primary driver of the outflow.

While convection naturally produces the apparently random localized
ejections, it does not by itself generate the observed magnetic-field
structure, nor do current models \citep{Freytag2017} produce the
necessary high velocities and hotspot temperatures. Finally,
interactions with a companion star or planet also do not directly
produce the observed stellar activity, directional ejections, and
magnetic morphology, even if engulfment can enhance a magnetic
dynamo. Instead, the observations appear to indicate a mechanism that
shares key characteristics with stellar and solar coronal mass
ejections (CMEs; \citealt{Webb2012}).

While these events differ from solar CMEs in temperature and thus
likely ionization state, they share the defining characteristics of
CMEs as impulsive, magnetically structured mass ejections: localized
magnetic footpoints, large-scale loop-like morphology
\citep{Webb2012}, and ballistic propagation following rapid
acceleration during launch \citep{Vourlidas2010,Chen2011}. In this
sense, the ejections from Mira A appear to represent a
low-temperature, weakly ionized analogue of slow solar CMEs that,
because of the size scales and masses involved, carry several orders
of magnitude more energy \citep{Chen2011,Emslie2012}. Typical
densities and measured magnetic-field strengths imply Alfv\'en speeds
comparable to the observed launch velocities, allowing magnetic
stresses to influence the dynamics during the eruption despite partial
ionization.

%======================================================================
\section{Conclusions}
\label{conclusion}
%======================================================================

The discovery of coronal-mass-ejection-style eruptions from an AGB
star redefines our understanding of late-stage stellar
evolution. These episodic magnetic ejections provide an additional
physical mechanism for the long-unexplained clumpiness, asymmetry, and
time variability of AGB mass loss, while directly seeding the complex
morphologies of post-AGB objects and planetary nebulae. Consequently,
magnetic activity must be integrated into models as a dynamically
dominant force in initiating mass loss and shaping circumstellar
environments. Given recent detections of directional mass ejections
from red supergiants (RSGs) \citep{Humphreys2007,Vlemmings2017VYCMa}
and the consensus that standard pulsations and convection fail to
support extended RSG atmospheres \citep{ArroyoTorres2015}, this
magnetic eruption mechanism may operate universally across evolved
cool luminous stars and contribute significantly to cosmic chemical
enrichment.

%======================================================================
% Figures
%======================================================================

\begin{acknowledgements}
WV, TK, MA, BBA, and LP acknowledge support from the
Olle Engkvist Foundation through grant 229-0368.
BL acknowledges funding from the European Union Horizon Europe
programme under the Marie Sk{\l}odowska-Curie grant agreement
No.~101126636.
MS acknowledges support from the Research Council of Norway,
ESGC project 335497.

This paper makes use of ALMA data
ADS/JAO.ALMA\#2024.1.01776 and
ADS/JAO.ALMA\#2011.0.00014.SV. ALMA is a partnership of ESO (representing its member states), NSF (USA) and NINS (Japan), together with NRC (Canada), NSTC and ASIAA (Taiwan), and KASI (Republic of Korea), in cooperation with the Republic of Chile. The Joint ALMA Observatory is operated by ESO, AUI/NRAO and NAOJ.
\end{acknowledgements}

\bibliographystyle{aa}
\bibliography{MiraCMe}

\begin{appendix}

\section{Table of maser spots}
  \begin{table*}
    \caption{Maser spots}
    \label{spots}
\begin{tabular}{rrrrrrr}
  id & $\delta$RA & $\delta$Dec & $V_{\rm lsr}$ & peak flux & $P_{\rm l, max}$ & EVPA \\
  & [mas] & [mas] & km~s$^{-1}$ & [Jy] &  & [$\circ$] \\
  \hline
  \hline
1.1 & $37.4 \pm 1.6$ & $-104.8 \pm 1.0$ & $47.21$ & $1.85$ & $0.73$ & $84.1 \pm 0.6$ \\
1.2 & $49.9 \pm 1.4$ & $-95.8 \pm 1.4$ & $46.95$ & $0.88$ & $0.67$ & $82.3 \pm 0.3$ \\
1.3 & $22.9 \pm 1.2$ & $-95.8 \pm 1.2$ & $47.56$ & $1.12$ & $0.59$ & $86.9 \pm 2.1$ \\
1.4 & $27.0 \pm 3.3$ & $-92.4 \pm 1.7$ & $47.38$ & $8.37$ & $0.49$ & $87.3 \pm 2.9$ \\
1.5 & $38.3 \pm 1.4$ & $-89.8 \pm 0.1$ & $47.04$ & $191.19$ & $0.75$ & $82.8 \pm 0.6$ \\
1.6 & $33.1 \pm 1.4$ & $-89.8 \pm 0.3$ & $47.82$ & $26.36$ & $0.64$ & $87.2 \pm 2.2$ \\
1.7 & $27.6 \pm 1.5$ & $-89.6 \pm 1.1$ & $46.92$ & $6.94$ & $0.72$ & $84.0 \pm 0.8$ \\
1.8 & $40.8 \pm 2.7$ & $-86.6 \pm 1.7$ & $47.02$ & $63.12$ & $0.71$ & $82.6 \pm 0.7$ \\
1.9 & $48.4 \pm 1.5$ & $-86.3 \pm 1.5$ & $46.74$ & $10.10$ & $0.67$ & $81.9 \pm 0.4$ \\
1.10 & $50.6 \pm 1.9$ & $-80.6 \pm 1.2$ & $47.12$ & $4.12$ & $0.76$ & $81.2 \pm 0.3$ \\
2.1 & $74.3 \pm 1.6$ & $-29.6 \pm 1.1$ & $42.43$ & $2.73$ & $0.56$ & $6.8 \pm 1.7$ \\
2.2 & $77.1 \pm 1.6$ & $-26.8 \pm 0.9$ & $42.86$ & $2.02$ & $0.55$ & $2.8 \pm 0.9$ \\
3.1 & $59.1 \pm 1.9$ & $8.3 \pm 1.6$ & $42.33$ & $5.35$ & $0.75$ & $-41.0 \pm 1.1$ \\
3.2 & $55.9 \pm 1.8$ & $15.2 \pm 1.8$ & $43.50$ & $0.54$ & $0.71$ & $-47.7 \pm 1.0$ \\
4.1 & $-16.1 \pm 1.4$ & $-6.6 \pm 2.0$ & $44.91$ & $0.86$ & $-$ & $-$ \\
4.2 & $4.9 \pm 1.1$ & $-5.8 \pm 1.1$ & $57.06$ & $1.39$ & $-$ & $-$ \\
4.3 & $1.9 \pm 0.8$ & $-5.8 \pm 0.8$ & $56.44$ & $2.48$ & $-$ & $-$ \\
4.4 & $-2.6 \pm 2.0$ & $-5.6 \pm 1.5$ & $55.01$ & $1.57$ & $0.30$ & $24.7 \pm 0.2$ \\
4.5 & $-6.8 \pm 1.4$ & $-3.1 \pm 1.4$ & $53.92$ & $2.09$ & $0.34$ & $23.7 \pm 0.9$ \\
4.6 & $-16.3 \pm 1.4$ & $-2.3 \pm 1.4$ & $52.74$ & $2.32$ & $0.40$ & $25.6 \pm 2.2$ \\
4.7 & $-11.8 \pm 1.6$ & $-0.0 \pm 1.6$ & $53.21$ & $8.76$ & $0.47$ & $22.9 \pm 0.3$ \\
5.1 & $2.4 \pm 1.5$ & $-92.6 \pm 1.5$ & $47.72$ & $1.50$ & $0.35$ & $-66.1 \pm 3.3$ \\
5.2 & $4.9 \pm 1.0$ & $-89.5 \pm 1.4$ & $46.19$ & $1.61$ & $0.31$ & $-64.2 \pm 1.1$ \\
5.3 & $2.6 \pm 1.7$ & $-86.8 \pm 1.3$ & $46.86$ & $0.96$ & $0.26$ & $-56.0 \pm 2.3$ \\
6.1 & $-40.1 \pm 1.3$ & $-20.8 \pm 1.3$ & $46.00$ & $0.98$ & $0.44$ & $20.0 \pm 2.5$ \\
7.1 & $-5.4 \pm 1.7$ & $15.7 \pm 1.4$ & $52.70$ & $2.61$ & $0.37$ & $35.3 \pm 4.7$ \\
8.1 & $-18.2 \pm 3.5$ & $16.0 \pm 1.4$ & $46.35$ & $14.60$ & $0.32$ & $40.5 \pm 1.4$ \\
9.1 & $-10.1 \pm 1.0$ & $26.4 \pm 1.6$ & $49.38$ & $1.84$ & $0.65$ & $53.3 \pm 0.3$ \\
9.2 & $-6.1 \pm 1.7$ & $26.8 \pm 1.5$ & $49.55$ & $6.34$ & $0.66$ & $55.1 \pm 2.6$ \\
9.3 & $-8.8 \pm 2.3$ & $30.3 \pm 1.2$ & $49.00$ & $7.97$ & $0.63$ & $52.8 \pm 0.6$ \\
10.1 & $1.9 \pm 1.1$ & $27.2 \pm 1.1$ & $42.81$ & $1.34$ & $0.58$ & $65.7 \pm 0.2$ \\
10.2 & $2.4 \pm 1.6$ & $30.2 \pm 0.7$ & $43.17$ & $3.86$ & $0.60$ & $65.8 \pm 0.3$ \\
10.3 & $7.9 \pm 0.9$ & $35.8 \pm 1.6$ & $50.08$ & $2.41$ & $0.56$ & $72.9 \pm 1.7$ \\
11.1 & $10.9 \pm 1.1$ & $36.2 \pm 1.1$ & $50.26$ & $1.49$ & $0.56$ & $73.3 \pm 1.8$ \\
11.2 & $13.9 \pm 1.0$ & $39.2 \pm 1.0$ & $49.49$ & $1.69$ & $0.45$ & $76.3 \pm 3.0$ \\
11.3 & $10.9 \pm 1.2$ & $39.2 \pm 1.2$ & $49.73$ & $1.24$ & $0.54$ & $75.1 \pm 3.1$ \\
11.4 & $17.6 \pm 1.6$ & $40.7 \pm 1.7$ & $49.22$ & $2.52$ & $0.48$ & $81.0 \pm 3.2$ \\
\hline
\multicolumn{7}{l}{Positions relative to Mira A at RA=$02h19m20.801s$ and Dec=$-02d58m45.625s$}
\end{tabular}
\end{table*}

\section{Functional form of the loop geometry}
\label{loop-model}

As described in \S~\ref{results}, the ad hoc adopted loop model, aimed
to be a simplified description of a CME, is a tube around a tilted
ellipse anchored to the stellar surface. We describe it in three
steps: the position vector $\mathbf r(\theta)$ tracing the ellipse's
central axis, the mapping of its line-of-sight component onto a
Doppler velocity, and the tube swept around this axis.

{\bf Central axis:} We work in a Cartesian frame $(x,y,z)$ centred on
the star, with $x=\Delta\mathrm{RA}$ and $y=\Delta\mathrm{Dec}$ the
sky-plane offsets and $z$ the line-of-sight coordinate (positive
towards the observer). The loop's shape is set by six free parameters:
the sky-plane footpoint centre $(x_0,y_0)$, the footpoint
half-separation $r_{fp}$, the position angle $\mathrm{PA}_0$ of the
chord between the two foorpoints, the semi-axes $H$ and $W$, and a
tilt angle $\phi$ that rotates the ellipse's plane about the
footpoint chord. The adopted values are listed in Table~\ref{tab:loop-params}.

Two footpoints are placed on the stellar surface (radius $R_\star$) by
offsetting from $(x_0,y_0)$ along the chord,

\begin{equation}
(x_{1,2},\,y_{1,2}) = (x_0,\,y_0) \pm r_{fp}\,(\sin\mathrm{PA}_0,\,\cos\mathrm{PA}_0),
\end{equation}

\noindent and lifting each onto the stellar sphere,
$\mathbf p_{1,2} = \big(x_{1,2},\,y_{1,2},\,\sqrt{R_\star^2-x_{1,2}^2-y_{1,2}^2}\big)$.
This anchors the loop in the south-west quadrant of the stellar disk.

From $\mathbf p_1$ and $\mathbf p_2$ we build a local orthonormal frame:
$\hat{\mathbf e}$ along the chord $\mathbf p_2-\mathbf p_1$, $\hat{\mathbf u}$
perpendicular to $\hat{\mathbf e}$ and lying in the plane containing the
local stellar radius vector, and $\hat{\mathbf v}=\hat{\mathbf e}\times\hat{\mathbf u}$
completing the right-handed triad. Defining $Y(\theta) \equiv H\sin\theta+y_c$
for $\theta\in[0,2\pi)$, the loop's central axis is the tilted ellipse

\begin{equation}
\mathbf r(\theta) \equiv \big(x(\theta),y(\theta),z(\theta)\big)
= \mathbf m + W\cos\theta\,\hat{\mathbf e}
+ Y(\theta)\big(\cos\phi\,\hat{\mathbf u}+\sin\phi\,\hat{\mathbf v}\big),
\label{eq:loop-axis}
\end{equation}

\noindent where $\mathbf m=\tfrac12(\mathbf p_1+\mathbf p_2)$ is the
footpoint midpoint, $\phi=0$ would leave the ellipse in the
$(\hat{\mathbf e},\hat{\mathbf u})$ plane containing the local stellar
radius, and $y_c$ is fixed by requiring the ellipse to pass exactly through
both footpoints, $y_c = H\sqrt{1-(2r_{fp}/2W)^2}$. Only the portion of the
curve with $|\mathbf r(\theta)|>R_\star$ is kept.

The projected position angle of the apex can be computed from Eq.~\eqref{eq:loop-axis}: writing
$\theta_{\rm apex}=\arg\max_\theta|\mathbf r(\theta)|$, it is the sky-plane
bearing of that point,

\begin{equation}
\mathrm{PA}_{\rm apex} = \arctan\!\big(x(\theta_{\rm apex}),\,y(\theta_{\rm apex})\big) \approx 95^\circ.
\label{eq:pa-apex}
\end{equation}

\begin{table}[h]
\centering
\caption{Adopted parameters of the loop's central axis (Eq.~\ref{eq:loop-axis}).}
\label{tab:loop-params}
\begin{tabular}{lcc}
\hline\hline
Parameter & Value & \\
\hline
$R_\star$            & $1.5$          & au \\
$(x_0,y_0)$          & $(-1.0,-0.3)$  & au \\
$r_{fp}$              & $0.15$         & au \\
$\mathrm{PA}_0$      & $5.4$          & deg \\
$H$ (semi-major)      & $7.5$          & au \\
$W$ (semi-minor)      & $7.0$          & au \\
$\phi$ (tilt)         & $-74.7$        & deg \\
\hline
\end{tabular}
\tablefoot{Full ellipse axes are $2H\approx15$~au and $2W\approx14$~au.}
\end{table}

{\bf Line-of-sight velocity:}
Since the axis along the line-of-sight for the observations corresponds to the velocity axis, we replace $z(\theta)$ for the line-of-sight component of $\mathbf r(\theta)$, with the projection of an outward expansion speed given by

\begin{equation}
V(r) = \left(\frac{r}{r_{\min}}\right)^{\beta},
\end{equation}

where $r=|\mathbf r(\theta)|$. This means that the line of sight velocity $V_{\rm LOS}$ corresponds to

\begin{equation}
V_{\rm LOS}(\theta) = V_{\rm *,lsr} - V(r)\,\frac{z(\theta)}{r}.
\label{eq:vlos}
\end{equation}

To match the observed velocity of the SiO masers, we use a power-law component $\beta=1.2$ for the radial velocity, which leads to a maximum radial velocity of $\sim16.4$~km~s$^{-1}$. This describes the current velocity field, which is the result of the interplay between gravity, magnetic forces and drag forces on the gas and, as described, would roughly correspond to a launch velocity of $\sim35$~km~s$^{-1}$.

{\bf Tube around the axis:}
The model is completed by constructing a tube of increasing size around the elliptocal loop. The tube radius grows from the footpoint to the apex as

\begin{equation}
a(r) = a_{\rm apex}\left(\frac{r}{r_{\rm apex}}\right)^{2},
\label{eq:tube-radius}
\end{equation}

with $a_{\rm apex}\approx1.5$~au chosen to reproduce the inferred
maser amplification path length near the apex, giving
$a(r_{\min})\approx0.015$~au at the stellar surface.
  
\end{appendix}

\end{document}